# Response to comment "Measurement of mobility in dual-gated MoS$_2$ transistors"

B. Radisavljevic, A. Kis[*]

[1]*Electrical Engineering Institute, Ecole Polytechnique Federale de Lausanne (EPFL), CH-1015 Lausanne, Switzerland*

*Correspondence should be addressed to: AndrasKis, andras.kis@epfl.ch

In our previous paper,[1] we report on switchable monolayer MoS$_2$ transistors with a high on-off ratio and we claim that dielectric screening can be used to increase the mobility of monolayer MoS$_2$. We estimate its mobility using a method previously applied by Lemme et al. to top-gated graphene nanoribbons[2], in which they extracted the channel mobility from the back-gating characteristic of the device, acquired with the top-gate disconnected.[2]

It is a common practice to extract the field-effect effective mobility using expression $\mu_{FE} = \left( dI_{ds}/d_{V_{gate}} \right) \times \left( L/WC_{ch,gate}V_i \right)$, where $L$ and $W$ are channel length and width, $C_{ch,gate}$ is the capacitance between the channel and gate per unit area and $V_i$ the voltage drop across the channel. It is common practice to make two approximations here: contact resistance is neglected, (equivalent to assuming $V_i = V_{ds}$ where $V_{ds}$ is the applied drain-source voltage, in practice, $V_i < V_{ds}$) and $C_{ch,gate}$ is assumed to be equal to the geometric capacitance between the gate electrode and the channel. For accurate measurements, both the contact resistance and the charge density (or corresponding capacitance) need to be measured.

Fuhrer and Hone suggest[3] that in our measurements, closely mirroring the approach previously used on top-gated graphene nanoribbons,[2] the back-gate and top-gate capacitance are equal. If this is indeed the case, then for all the devices presented in the manuscript we would be systematically underestimating $C_{bg}$ by a factor of 14 and not ~43 as Fuhrer and Hone claim, because we are only covering one third of the channel with the top gate and HfO$_2$ dielectric ($k_{HfO2}$=19). For the device presented in the manuscript, this would reduce the mobility estimate from ~ 217 to 15 cm$^2$/Vs, and for the first device in table 1 from the supplementary material[1] from 780 cm$^2$/Vs to 55 cm$^2$/Vs. This is in fact one and the same device which deteriorated between the first measurement in August 2010 and the second set of measurements in October 2010 solicited by the referees. The two-contact effective mobility for this device without the dielectric was 0.2 cm$^2$/Vs. Therefore, even if the proposed mistake in the capacitive coupling was made, the resulting numbers would still be significantly higher than the only previous report on monolayer MoS$_2$ (3 cm$^2$/Vs)[4], and there would remain a factor 275 increase in effective mobility following the deposition of the dielectric. However, we cannot make an accurate estimate while neglecting the contact resistance. In a previous paper on a similar WSe$_2$ transistors, a similar material, a factor of 5 error due to neglecting the contact resistance was reported,[5] meaning that the value of 55 cm$^2$/Vs could be an underestimate of the actual mobility in the device by a similar factor.

Overlapping back-gating and top-gating characteristics (fig b in the comment) do not prove that the capacitances are equal in both cases, they only prove that the $C_{ch,gate} \cdot V_i$ product is equal. $V_i$ would be equal in both cases only if the contact resistances would be the same. They are not: in the bottom-gating case, our contact resistance is reduced because the entire channel is gated ($V_i$ closer to $V_{ds}$ = 10mV), while in the top-gated case one third of the channel is gated while two thirds of the channel act as contacts. Consequently, the access resistance is increased (smaller actual $V_i$). Similarly, one cannot estimate the mobility from the case of $V_{tg}$ = 0 either because only one third of the channel is covered by the top gate. The model proposed by Fuhrer and Hone therefore cannot be used to extract the correct value of the mobility from two-contact measurements.

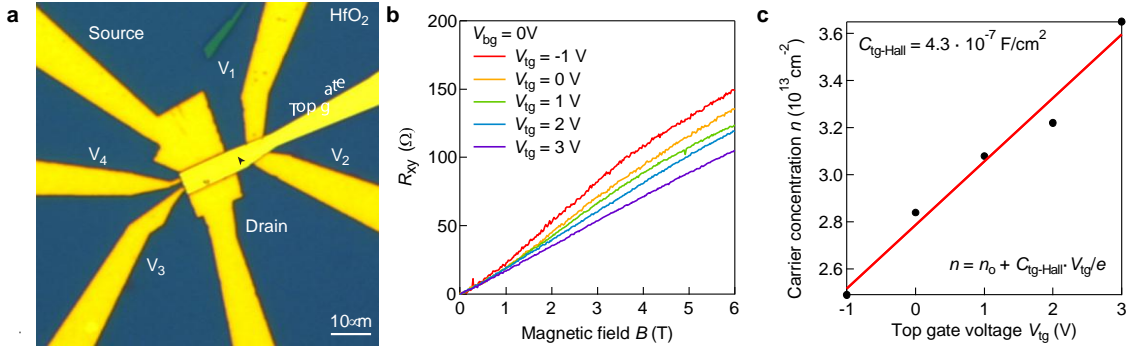

**Figure 1. Hall effect measurements on monolayer MoS$_2$. a** Optical image of a Hall-bar device based on monolayer MoS$_2$ with a top gate and HfO$_2$ top-gate dielectric. **b** Hall-resistance of the device as a function of the magnetic field, recorded for different values of the top-gate voltage $V_{tg}$. **c** Charge carrier concentration extracted from Hall-resistance and its dependence on the applied top-gate voltage. The linear fit allows the top-gate capacitance to be extracted.

Being aware of the inherent limitations of two-contact, we have also characterized monolayer MoS$_2$ using Hall-effect measurements where we can measure the mobility, contact resistance and capacitive coupling simultaneously figure 1. (ref 6). We present here the main findings relevant to this correspondence. Devices without a top-gate dielectric show a mobility increasing with temperature, behavior indicative of charged-impurity scattering, followed by decreasing mobility with temperature due to electron-phonon scattering. Top-gated devices show a temperature-independent higher mobility in the low-T region, while in the high-T region, the mobility decays slower than expected ($\mu \sim T^{-a}$, with a < 1). The mobility of one of these devices varies from 168 cm$^2$/Vs at 4 K to 60 cm$^2$/Vs at 250 K. This shows that the deposition of the top-gate dielectric enhances the device mobility and that this enhancement is consistent with suppression of Coulomb scattering and a modified electron-phonon scattering, as proposed in the original manuscript.[1]

From the Hall-effect measurements we also measure the capacitive coupling between the channel and the gate and find that the capacitive coupling can be increased by a factor of 3 following only the deposition of the top-gate dielectric and a factor of 50 for the floating gate/dielectric case. In this case, the top gate covered the entire channel. We also find that neglecting the contact resistance underestimates the mobility by a factor of ~2.5. While these measurements prove that the capacitance can be underestimated in a complicated dielectric environment, the explanation proposed by Fuhrer and Hone would give identical results for the bare and covered $MoS_2$ in the absence of the bonding pad or a top-gate electrode.

To conclude, measurements based on the Hall-effect show that the practice of assuming that the channel-gate capacitance does not change when the dielectric environment changes is wrong. However, because of conflicting influences of the underestimated capacitance and neglected contact resistance, we cannot make a more precise mobility estimate based on our previous two-contact measurements alone and multiplying mobility values by correction factors that are not based on actual measurements on the same device would not result in better estimates. Measurements where we can accurately measure both the contact resistance and capacitive coupling show mobility in monolayer $MoS_2$ that is significantly higher than the only previous report on monolayer $MoS_2$ (Ref 4) and show a clear enhancement of mobility with dielectric deposition. One of the main conclusions of our previous paper, that $HfO_2$ deposition can be used to increase the mobility of monolayer $MoS_2$ up to values comparable to thin-film silicon is therefore still valid. Other device characteristics presented in the paper such as the fact that a monolayer $MoS_2$ transistor can be turned off, can have a $10^8$ current on/off ratio and negligible leakage current are independent of the actual value of the extracted charge carrier mobility.